# William H. Kruskal, Mentor and Friend

**Judith M. Tanur**

Toward the end of 1963 I was interviewed by David Sills who was Editor in Chief of the *International Encyclopedia of the Social Sciences*, then in preparation. The job opening was for a staff editor to assist Bill Kruskal in his work as associate editor for statistics for the *Encyclopedia*. I later learned (by judicious snooping into personnel files) that after the interview David informed Bill that "Mrs. Tanur is young and pliable." It is always interesting to explicate the exact stimulus that gives rise to a response—in this case I strongly suspect that it was my reply to a particular question posed during the interview that caused David to make that judgment (though I was indeed quite young at the time). David, having had to referee entirely too many intellectual battles between associate editors and staff editors, had asked me if I would be willing to take guidance from Bill. My genuine shock at the thought that I might not be willing to—I think I stammered something about his being William Kruskal, one of the originators of the famous Kruskal–Wallis test and me being a recent MA in mathematical statistics and so of course I'd be willing to take guidance—probably contributed heavily to my being hired.

How lucky I was to be hired! The job introduced me to many of the greats in contemporary statistics and launched me into a career of editing and explicating statistics. Most importantly, it gave me my first chance to work with Bill. What a role model! In his work on the *Encyclopedia*, Bill cared about everything—and I mean everything—from weighty issues of content and exposition to tiny issues of typography: was there really a two-point space between a symbol and its subscript? I took a self-taught crash course in the printing of mathematics, but learned whatever I know of substantive statistics by working with Bill and the contributors as we edited, asked for rewriting and edited again. When we asked for a rewrite, often Bill would do a sample to show the contributor what he had in mind, but he was always very careful not to take over the article, including in each such mailing the disclaimer that he didn't want to put words into the contributor's pen. Nevertheless, many of the contributors found it most convenient to just adopt Bill's sample rewriting. Thus, much of the material both in the *International Encyclopedia of the Social Sciences* and in its offshoot, the *International Encyclopedia of Statistics* that Bill and I edited, came from Bill's pen. Indeed, much more than has ever been acknowledged.

*Judith M. Tanur is Distinguished Teaching Professor Emerita, Department of Sociology, State University of New York, Stony Brook, New York 11794-4356, USA e-mail: jtanur@notes.cc.sunysb.edu.*

More broadly, the coverage of statistics in the encyclopedias represents Bill's conceptualization of the field and its ramifications. Well before I joined the staff he had been the leader of the project that decided what articles to include, and the decisions were indeed encyclopedic. Right next to articles on Estimation (separate articles for Point Estimation and Confidence Intervals and Regions) there are articles on Errors (separate articles on Nonsampling Errors and on The Effects of Errors in Statistical Assumptions). As well as several articles on Time Series and on Index Numbers, there are several on Nonparametric Statistics and on Multivariate Statistics. The system of grouping articles together implicit in these titles and an elaborate system of cross-referencing makes the mapping of the field both a fascinating glimpse into Bill's own mental representation of statistics writ large and a useful guide to a student trying to connect parts of the field. I was lucky enough to be the first such student to have as an assignment reading these encyclopedia articles—and reading them in revision, and reading them after copyediting, and reading them in galley, and reading them in page proofs!

Bill's consideration for me while we were in day to day contact working on the *Encyclopedia* was enormous. In professional matters, because I was so "young and pliable" and indeed so much junior, Bill could easily have used me as a glorified secretary, but he insisted on treating me as a full partner in the enterprise, often asking my advice and consent at times when it would have been much easier to just go ahead with his own judgment. In addition, because I had to take responsibilities that I often felt were beyond my abilities, I learned both statistics and diplomacy.

Bill's consideration for me also extended to personal matters: Let me illustrate with a story about a minor point instead of a more dramatic one. I remember Bill's visiting my little cubbyhole of an office at Free Press and asking if it would be all right to smoke one of the little cigars he favored at that time. I said, "Of course," and hoping to put him at his ease somewhat more, added that my husband also smoked cigars. "Yes," said Bill, "but that's why you get out of the house to work, isn't it?"

Bill and I continued to work together after the *International Encyclopedia of the Social Sciences* appeared. We spent five years negotiating a contract with Free Press that would permit us to publish the statistics articles from that *Encyclopedia* as a separate *International Encyclopedia of Statistics*. Then we worked another five years to update and correct the articles and solicit several new articles to fill holes in the original social science encyclopedia. The contract took so long to negotiate because Bill's perfectionism had rubbed off on me: together we insisted on having control of every step of the production process. Thus we worked on the system of signaling what material was new, on checking the bibliography, and supervised and ended up constructing the index ourselves—proving to Bill once and for all that every volume can have a fine index. Many of us learned that lesson from Bill; the volumes we write and edit are the richer for it and our readers owe a debt of gratitude to Bill for their convenience in using such volumes.

It was at Bill's suggestion that Fred Mosteller invited me to become part of the team that produced *Statistics: A Guide to the Unknown*—another wonderful learning experience for me. Again I learned more about statistics, learned more about expository skills and gained more admiration for my co-editors. There were those who said that a readable volume about statistics for the lay public couldn't be done. Gentle persuasion of authors, good examples from their own pens and extreme persistence, especially from Fred and Bill, made it happen. And the volume has a wonderful index. A new fourth edition, edited by a completely new and younger team, has just come out.

I did not become a member of the Committee on National Statistics until some time after Bill had served as the Committee's first chairman, but I knew of Bill's dedication to the integrity of federal statistics through articles we had edited and through copies of his correspondence that he had forwarded to me over the years. In some sense that exposure had been Bill's way of giving me some basic training in the importance of federal statistics and the workings of federal statistical agencies. My service on the Committee served as a postgraduate course in these matters. Indeed, using service on the Committee—either by membership or on staff—to educate neophytes about the importance of federal statistics was an important part of Bill's agenda. This is yet another mode of Bill's teaching and one that continues to be emulated by the Committee on National Statistics.

Those of us who knew and loved Bill were on his mailing lists, and so were our friends and colleagues, and so were their friends and colleagues. . . . Because he was a true polymath, Bill's reading was broad and eclectic, and he was able to make surprising connections between and among people and ideas. So, as we all know, when he found something that might interest one or more of his pen pals, he photocopied and mailed. Over the years I was amused, edified, delighted and awed by this breadth of interests. Those mailings stopped some years ago—and I have missed them. I have often thought how delighted Bill would have been if he had been young enough to take advantage of the electronic wonders of scanning and e-mail. Then, however, our electronic mailboxes would have been as overwhelmed as our paper files indeed were by the volume of Bill's correspondence. Like our files, our minds are fuller and better furnished for having had Bill Kruskal in our lives.